\documentclass[lettersize,journal]{IEEEtran}
\usepackage{amsmath,amsfonts}
\usepackage{algorithmic}
\usepackage{algorithm}
\usepackage{array}
\usepackage[caption=false,font=normalsize,labelfont=sf,textfont=sf]{subfig}
\usepackage{textcomp}
\usepackage{stfloats}
\usepackage{url}
\usepackage{verbatim}
\usepackage{graphicx}
\usepackage{epstopdf}
\usepackage{cite}
\usepackage{hyperref}
\usepackage{pifont}
\usepackage{bbm}

\usepackage{multicol}
\usepackage{multirow}
\begin{document}


\title{Uncertainty Quantification in Melody Estimation using Histogram Representation{\thanks{This work was supported by grant no. PB/EE/2021128B from Prasar Bharti.}}}

\author{Kavya Ranjan Saxena,~\IEEEmembership{Graduate Member,~IEEE}, Vipul Arora,~\IEEEmembership{Member,~IEEE}\\{kavyars@iitk.ac.in, vipular@iitk.ac.in}\\{Department of Electrical Engineering}\\{Indian Institute of Technology Kanpur, India}}

\markboth{Journal of \LaTeX\ Class Files,~Vol.~14, No.~8, August~2021}%
{Shell \MakeLowercase{\textit{et al.}}: A Sample Article Using IEEEtran.cls for IEEE Journals}

\maketitle 
\begin{abstract}

Confidence estimation can improve the reliability of melody estimation by indicating which predictions are likely incorrect. The existing classification-based approach provides confidence for predicted pitch classes but fails to capture the magnitude of deviation from the ground truth. To address this limitation, we reformulate melody estimation as a regression problem and propose a novel approach to estimate uncertainty directly from the histogram representation of the pitch values, which correlates well with the deviation between the prediction and the ground-truth. We design three methods to model pitch on a continuous support range of histogram, which introduces the challenge of handling the discontinuity of unvoiced from the voiced pitch values. The first two methods address the abrupt discontinuity by mapping the pitch values to a continuous range, while the third adopts a fully Bayesian formulation, which models voicing detection as a classification and voiced pitch estimation as a regression task. Experimental results demonstrate that regression-based formulations yield more reliable uncertainty estimates compared to classification-based approaches in identifying incorrect pitch predictions. Comparing the proposed methods with a state-of-the-art regression model, it is observed that the Bayesian method performs the best at estimating both the melody and its associated uncertainty.

\end{abstract}

\begin{IEEEkeywords}
melody estimation, uncertainty estimation, histogram loss, Bayesian models 

\end{IEEEkeywords}

\section{Introduction}
The fundamental task in the field of music information retrieval is to estimate singing melody from polyphonic audio, which has applications in downstream tasks such as music recommendation \cite{musicrecom}, cover song identification \cite{coversong}, music generation \cite{musicgen}, and voice separation \cite{voicesep}.

Previous machine-learning approaches for melody estimation from polyphonic audio have typically formulated this problem either as a multi-class classification~\cite{nmf-crnn}~\cite{saxena2024} or as a time-frequency segmentation problem~\cite{deepsalience}~\cite{patch-basedcnn}. In the classification-based methods, the continuous pitch range is discretized into pitch classes. Given a spectrogram representation of an audio, each time frame is assigned to one of the pitch classes. The target for each time frame is represented as a one-hot vector or a Gaussian-blurred version to reduce sensitivity to small pitch deviations. The models in this setup are trained using categorical cross-entropy loss. An alternate method~\cite{jdc} is to decompose melody estimation into voicing detection and pitch estimation, where both are treated as classification problems. In contrast, segmentation-based methods define the target as a two-dimensional salience map, where each column corresponds to a time frame and each row to a frequency bin. At each time frame, the target column is obtained by quantizing the pitch to the nearest frequency bin $f$ and applying a Gaussian blur centered at $f$, assigning non-zero weights to neighboring bins. This soft labeling reduces sensitivity to small pitch deviations. The models in this setup typically minimize a binary cross-entropy loss between the predicted and target salience map. Another hybrid method~\cite{en-decoder} treats voicing detection as a classification problem and pitch estimation as a segmentation problem.

While these methods can also provide confidence scores from softmax probabilities in classification-based methods~\cite{saxena2024} or peak salience values in segmentation-based methods, they only reflect the model certainty and do not capture the magnitude of pitch deviations. For instance, a prediction that is one semitone away from the ground truth is penalized equally as a prediction that is several semitones away, provided both predictions are wrong. Consequently, the confidence score does not convey how close the prediction is to the correct pitch in continuous frequency space. This limitation motivates formulating melody estimation as a regression problem, where the models predict continuous-valued pitch so that uncertainty can be quantified directly in terms of pitch deviations.

Previous regression-based approaches have primarily been applied to monophonic speech data rather than polyphonic music. For instance, some hybrid methods~\cite{nonstationaryspeech}~\cite{directf0est} treat voicing detection as a classification problem and pitch detection as a regression problem trained using mean squared error (MSE) loss. While MSE predicts point estimates for each time frame, it implicitly assumes a constant variance, treating all prediction errors equally. This means that the model cannot express the uncertainty about a particular prediction. A more flexible approach is probabilistic regression, where the target pitch is modeled as a Gaussian random variable whose mean and variance depend on the input. For example, a state-of-the-art method~\cite{seitzer} for estimating uncertainty in regression assumes that given a sample $(x,y)$, the target $y$ is conditionally dependent on input $x$ and follows a normal distribution $\mathcal{N}(\mu(x),\sigma^2(x))$. The estimates $\hat{\mu}(x)$ and $\hat{\sigma}^{2}(x)$ of the true mean and variance are estimated by training the model using negative log-likelihood loss. The estimated variance $\hat{\sigma}^2(x)$ represents the uncertainty that varies with input $x$. However, this Gaussian assumption has limitations, such as it assumes a unimodal symmetric distribution, and struggles to capture complex multimodal patterns in the data. An alternative is the histogram-based regression~\cite{histogram_loss} that models pitch as a continuous random variable by approximating its distribution with a histogram. Instead of assuming a single Gaussian, the model predicts a full probability distribution over the support pitch range. This representation accommodates multi-modal patterns in the data from which mean and variance can be directly computed, making it a more expressive way to model uncertainty in pitch estimation. 

In this paper, we approach melody estimation as a regression problem that explicitly focuses on predicting uncertainty correlated with pitch deviations. We design three methods using histogram representations to model pitch, requiring the support range to be continuous. In the first two methods, the discontinuity between unvoiced and voiced frequency ranges is handled by transforming them into a continuous range. Given a spectrogram as an input, the model predicts a distribution over this continuous range for both unvoiced and voiced time frames. The third method adopts a Bayesian framework, treating voicing detection as classification and voiced pitch estimation as a regression problem. In this case, given a spectrogram as an input, the model classifies the unvoiced and voiced time frames and simultaneously predicts the distribution only for the voiced time frames. The proposed uncertainty estimation method ensures that larger prediction errors correspond to higher uncertainty, and smaller errors to lower uncertainty. A point to note here is that the uncertainty estimates in the first two methods are obtained for both unvoiced and voiced frames, whereas in the third method, they are obtained only for the voiced frames.

The main contributions of this work are:
\begin{itemize}
    \item Treating melody estimation from polyphonic audio as a histogram-based regression problem with continuous pitch prediction. To the best of our knowledge, there are no deep models yet that do so.
    \item A novel method to estimate uncertainty from histogram representation by maximizing the likelihood of prediction. This uncertainty correlates with the deviation of the mean of the estimated distribution from the ground truth.
    \item Experimental comparison of the performance of proposed methods against state-of-the-art models. 
\end{itemize}

The codes of the proposed models are available online at~\href{https://github.com/KavyaRSaxena/me_reg_taslp}{https://github.com/KavyaRSaxena/me\_reg\_taslp}.

\section{Related Works}
\label{relatedworks}

\subsection{Existing works on melody estimation}
Existing work on extracting pitch from monophonic audio includes CREPE~\cite{crepe}, which predicts pitch directly from the time-domain audio. Another monophonic method that can also extend to a polyphonic method estimates pitch with uncertainty, such as SPICE~\cite{spice}, which combines a confidence head for voicing and a regression head that represents pitch in a latent, continuous space that is linearly related to semitones. Further, various neural network-based methods have been proposed to extract melody from polyphonic audio. For instance, Lu et al.~\cite{aud-sym-tl} use a DCNN with dilated convolutions for semantic segmentation of candidate pitch contours, while Bittner et al.~\cite{deepsalience} employ a fully convolutional network to learn salience representations for fundamental frequency estimation. Encoder-decoder architectures~\cite{en-decoder} improve performance by separately modeling voiced and unvoiced frames. Other strategies include joint voicing detection via classification~\cite{jdc}, attention networks~\cite{attention} for capturing frequency relationships, and semi-supervised or knowledge distillation frameworks such as HKDSME~\cite{hkdsme}, MTANET~\cite{mtanet}, and HANet~\cite{hanet} to capture harmonic structures and long-range dependencies. The performance of the melody estimation model can be further improved by performing domain adaptation~\cite{mlsp_kavya}.  

All the above deep-learning methods treat melody estimation as a multi-class classification or segmentation task.

\subsection{Existing works on uncertainty in regression}
In regression, by assuming that the target follows a particular distribution, the model is trained by minimizing the negative log-likelihood~\cite{nll}, ensuring that the predicted mean and variance closely match the true data distribution. The model variance captures the uncertainty of the prediction. One such work~\cite{kendall} uses Monte-Carlo Dropout~\cite{montecarlo} to sample multiple predictions by applying different dropouts, allowing the empirical distribution of these predictions to capture the predictive uncertainty. Similarly, another work~\cite{lakshminarayanan} achieves the same goal by using an ensemble of models, where predictions from multiple independently trained models are aggregated to estimate the uncertainty. There are other works~\cite{chua}~\cite{seitzer} that also focus on capturing the predictive uncertainty. However, a key limitation of these models is that they often produce overconfident variance estimates~\cite{seitzer}, which are addressed by some methods~\cite{sol1}~\cite{sol2}.

\section{Preliminaries}
\label{prelim}

\subsection{Histogram Loss}\label{hist-loss}
The regression problems commonly involve minimizing mean squared error loss or L2 loss. This is analogous to the maximum likelihood estimation of the output modeled as a Gaussian random variable with a fixed variance. The final prediction is the mean of this distribution. Instead of computing a point estimate, the histogram loss~\cite{histogram_loss} (denoted by HL) computes a density function that improves the generalizing capability of the model by capturing the entire distribution of possible outcomes, rather than a single point estimate. This representation allows the model to better account for uncertainties and variabilities in the data, leading to more accurate and reliable predictions. 

Consider a sample $(x,y)$, where $y$ is a continuous target corresponding to some input $x$. Instead of directly predicting $y$, we select a target distribution on $y|x$. Suppose this target distribution has a support range $[a,b]$, pdf $p$, and CDF $F$. Our goal is to learn the parameterized predictive distribution $q(y|x)$ by minimizing the KL divergence to $p$. We restrict the predictive distribution $q(y|x)$ to be a histogram density, where the support range $[a,b]$ is uniformly partitioned into $K$ bins of equal width $b_w=\frac{b-a}{K}$. 

Consider a model $f_{\theta}$ parameterized by $\theta$ that predicts the bin probabilities. The predictive distribution is given by $q(y|x) = f_{\theta}(x) = (q_{1},q_{2},...,q_{K}); \quad k=1,2,...,K$,
where $q_{k}$ represents the probability that $y$ falls within the $k^{th}$ bin, i.e., $q_{k} = P(y \in [l_k,l_k+b_w]|x)$, with left bin edge as $l_k = a + (k-1)b_w$. By construction, the predicted bin probabilities satisfy $\sum_{k=1}^{K}q_{k}=1$. The KL divergence between $p$ and $q$, given as:
\begin{equation}
    KL_x(p||q) = H_X(p,q) - H_X(p)
\end{equation}
where  $H_x(p,q)$ is the cross-entropy between $p$ and $q$ and $H_x(p)$ is the entropy of $p$. Since $H_x(p)$ is constant with respect to the model parameters, minimizing the KL divergence reduces to minimizing the cross-entropy:
\begin{equation}
    \begin{aligned}
        H_x(p, q) &= -\int_a^b p(y) \log q(y) \, dy \\
        &= -\sum_{k=1}^{K} \int_{l_k}^{l_k + b_w} p(y) \log q_{k} \, dy \\
        &= -\sum_{k=1}^{K} \log q_{k} \underbrace{\left( F(l_k + b_w) - F(l_k) \right)}_{p_{k}} \\
    \end{aligned}
\end{equation}
Therefore, this gives the histogram loss as:
\begin{equation}
    HL_x(p,q) = - \sum_{k=1}^K p_{k} \log q_{k}
\label{hil}
\end{equation}
where $p_k$ is called as the bin weights. The choice of target distribution $p$ is flexible as long as its CDF $F$ can be evaluated for each bin $k$. In this work, we consider a Gaussian distribution as the target distribution. Notably, since the target distribution is fixed, the bin weights $p_{k} =  F(l_k + b_w) - F(l_k)$ can be precomputed for each sample, making model training computationally efficient. An important benefit of the histogram loss is that the divergence between the predictive distribution and the full target distribution $p$ can be computed very efficiently. In addition, selecting a different form of the target distribution merely changes the weighting terms in the cross-entropy. A point to note is that for histogram loss to be applicable, the support range $[a,b]$ must be continuous and uniformly partitioned. 

\subsection{Uncertainty in Regression}\label{nll-loss}
Consider a sample $(x,y)$. Assuming that the target follows a particular distribution conditioned on $x$, i.e. $y|x$, we consider a model $f_{\theta}$, parameterized by $\theta$ which outputs a predictive distribution as $q(y|x) = \mathcal{N}(\hat{\mu}(x),\hat{\sigma}^{2}(x))$. With $x$ as an input, the model predicts $f_{\theta}(x) = [\hat{\mu}(x),\hat{s}(x)]$, where $\hat{\mu}(x)$ is the predicted mean of the target and $\hat{s}(x)$ is the log-variance. The predicted variance can be calculated as $\hat{\sigma}^{2}(x) = \text{exp}(\hat{s}(x))$ which captures the uncertainty in the model prediction~\cite{seitzer}.

The parameters $\theta$ of the model are trained using negative log-likelihood loss $\mathcal{L}_{NLL}$ defined as:
\begin{equation}
    \mathcal{L}_{NLL} = -\mathbb{E}_{x,y} \Bigg[\frac{1}{2} \log \hat{\sigma}^{2}(x) + \frac{(y-\hat{\mu}(x))^2}{2\hat{\sigma}^2(x)} + \text{const} \Bigg]
\end{equation}

\section{Methodology} 
\label{method}
The audios are merged into a single channel and downsampled to 16kHz. Since the duration of the audios may be different, we have divided the audios into chunks of 1-second each. We calculate the spectrogram $X$ of dimension $M\times T$ of the audio chunks using a short-time Fourier transform.  The spectrogram is calculated using a 2048-point Hann window and a hop size of 10ms, where $M=1025$ is the number of frequency bins and $T=100$ is the number of time frames.

\subsection{Data Preparation}
Consider a sample $(X,y)$. Let the input be a spectrogram $X\in \mathbb{R}^{M\times T}$, where $M$ is the number of frequency bins, and $T$ is the number of time frames. The output $y$ is a vector of dimension $T$ consisting of frequency values (in Hz) corresponding to each time frame $t$. The frequency value $y_t$ at each time frame $t$ can either be unvoiced (represented as 0) or voiced, with voiced frequencies ranging from $51.91$ Hz $(G\#1)$ to $830.61$ Hz $(G\#5)$ with a resolution of 1/8 semitone, i.e., $B=96$ bins per octave. The voiced frequency range is non-uniformly spaced, following a geometric progression where each semitone corresponds to a frequency ratio of $2^{(1/B)}$ relative to the previous one. Moreover, the support of $y_t$ is discontinuous, since it includes the unvoiced value 0 and the voiced frequency range, i.e. $\{0\} \cup [51.91,830.61]$. Due to this discontinuous and non-uniform support, the histogram loss cannot be applied directly. To address this, we transform the output $y$ for each voiced time frame $t$ into log-frequency values, calculated as

\begin{equation}
    g(y_t) = \log_2 \Bigg( \frac{y_t}{51.91} \Bigg)
\label{log_freq_fn}
\end{equation}
where $51.91$ Hz represents the lower bound of the voiced frequency range under consideration. Applying the transformation as in eq.~\ref{log_freq_fn}, the log-frequency values for voiced frames are restricted to the voiced support range $[0,4]$, where $g(51.91) = 0$ and $g(830.61) = 4$. The voiced support range is discretized with a uniform bin width $b_w=0.01042$. Since unvoiced frames are not transformed, the support range still remains discontinuous, preventing direct application of histogram loss. To address this, we propose methods to handle unvoiced frames and create a continuous and uniform support range, as described below.


\subsection{Histogram loss with fixed standard deviation \texorpdfstring{$\sigma$}{sigma} (M1)} \label{M1}
For a sample $(X,y)$, the frequency value at each unvoiced frame $t$ of the output $y$ is mapped to a bin that is uniformly $50$ bins below $g(51.91)$, i.e., $g(51.91)-(50\times b_w) = -0.521$. 
We choose a value of 50 bins to replicate or maintain a sufficient gap between unvoiced and voiced log-frequency values, at the same time keeping in mind the computational complexity, as it increases with the increasing number of uniform bins. With this modification, the original discontinuous and non-uniform support range $\{0\} \cup [51.91,830.61]$ Hz is now transformed into a continuous and uniformly partitioned range $[-0.521,4]$ in log scale, resulting in a total of $K=435$ uniformly partitioned bins. Here, $k=1$ represents the unvoiced bin and $k\in [k_{v1},k_{v2}]$ represents the voiced bins, where $k_{v1} = 51$ and $k_{v2}=435$.

Consider a dataset $D = \{(X_i,y_i)\}_{i=1}^{I}$, where $X_i$ is the spectrogram of shape $M\times T$, and $y_i$ is a vector of dimension $T$, consisting of log-frequency values for voiced frames computed using eq.~\ref{log_freq_fn}, with unvoiced frames mapped to $-0.521$. Each time frame $t$ of the $i^{th}$ sample is either classified as voiced or unvoiced, i.e., $c \in \{0,1\}$. The weights $w_c$ for each class are calculated as:

\begin{equation}
    w_c =
    \begin{cases} 
        \frac{\sum_{i,t} \mathbbm{1} \{c_{it}=1\}}{\sum_{i,t} 1}, & \text{if $c$ = 1} \\ 
        1 - w_1, & \text{if $c$ = 0}
    \end{cases}
    \label{wcs}
\end{equation}
where $c_{it}=1$ denotes the voiced time frame $t$ and $c_{it}=0$ denotes the unvoiced time frame $t$, for the $i^{th}$ sample. For simplicity, we ignore the sample index $i$ while explaining the notations.  

Consider a sample $(X,y)$. For a particular time frame $t$, we consider a target distribution $p(y_{t}|X)$ as a Gaussian distribution within a support range $[-0.521,4]$, with mean $y_{t}$ and standard deviation $\sigma_t$ equal to bin width $b_w$, i.e., $p(y_{t}|X)=\mathcal{N}(y_{t},b^2_w)$. The bin weight $p_{tk} = F(l_k+b_w) - F(l_k)$ for each bin $k$ is already computed offline, making $p_{t}$ of dimension $K$.

As a result, the dataset is reformulated as $D = \{(X_i,y_i,p_i)\}_{i=1}^I$, where $p_i$ represents the bin weights of dimension $K\times T$. For simplicity, we consider a single sample $(X,y,p)$. Consider a base model $f_{\theta}$, where $\theta$ are the model parameters. For a particular time frame $t$, the base model $f_{\theta}$ predicts the predictive distribution $q(y_{t}|X)$ which consists of predicted bin probabilities $(q_{t1},q_{t2},...,q_{tK})$ of dimension $K$. 
During training, the parameters $\theta$ are updated using the gradient descent algorithm as,
\begin{equation}
    \theta \leftarrow \theta - \alpha \nabla_{\theta}\mathcal{L}_{wHL}(f_{\theta})
    \label{gd}
\end{equation}
where $\alpha \in \mathbb{R}^+$ is the learning rate, and $\mathcal{L}_{wHL}$ is the weighted histogram loss defined as:
\begin{equation}\label{whl}
\begin{gathered}
    \mathcal{L}_{wHL} =  -\sum_{i,t} w_{c_{it}} \sum_{k=1}^{K} p_{itk} \log q_{itk}    
\end{gathered}
\end{equation} 
where weights $w_{c_{it}}$ denote the class weight corresponding to the voiced or unvoiced time frame $t$ for $i^{(th)}$ sample ($c_{it}=1$ or $c_{it}=0$), obtained from the eq.~\ref{wcs}.
After training the base model $f_{\theta}$ for $E_1$ epochs, the mean of the predicted distribution at time frame $t$ is given by 
\begin{equation}
    \hat{y}_{t} = \mathbb{E}_{\hat{y}\sim q(y_{t}|X)}[\hat{y}]
    \label{eq:expected_y}
\end{equation} 

\begin{figure}
    \centering
    \includegraphics[width=\columnwidth]{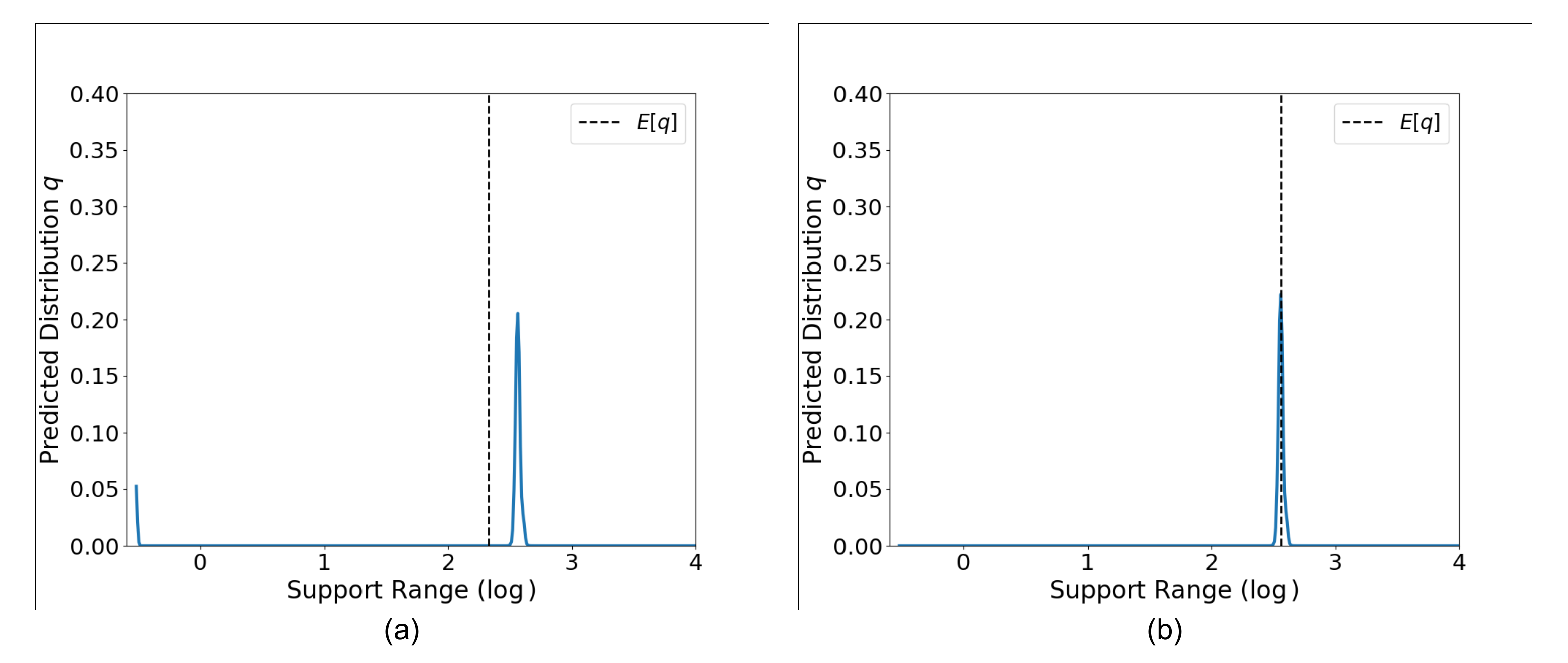}
    \caption{Predicted distribution $q(y_t|X)$ at a particular time frame $t$ with (a) two simultaneous peaks at unvoiced and voiced bins (incorrect point estimate), and (b) updated distribution after applying the pruning algorithm (correct point estimate).}
    \label{prune-example}
\end{figure}

\begin{algorithm}[t]
  \small
  \caption{Pruning Algorithm $P$}
  \label{pruning}
  \begin{algorithmic}[1]
    \REQUIRE Trained model $f_{\theta}$
    \REQUIRE $\delta = 0.01 $ (probability threshold); $\Delta k = 10$ (number of bins to suppress around the selected peak)
    \REQUIRE Sample $(X,y)$ and predicted distribution $q(y|X)$ of dimension $K\times T$
    \FOR {$t$ frames in $X$}
        \STATE Obtain $q(y_t|X) = (q_{t1},q_{t2},...,q_{tK})$ 
        \IF {$q_{t1} \geq \delta$ and $\max_{k \in [k_{v1},k_{v2}]} q_{tk} \geq \delta$}
            \STATE Select bins where unvoiced and voiced peaks are present, i.e., $k_{uv} = 1$ and $k_v = \underset{k \in [k_{v1},k_{v2}]}{\arg\max} \, q_{tk}$ 
        \STATE Select the bins to suppress the probability values, i.e, \\
            $k_{sup} = \begin{cases}
                \{k_{uv},..,k_{uv}+\Delta k\} & \text{if } q_{tk_{uv}} < q_{tk_{v}} \\
                \{k_{v}-\Delta k,..,k_v,...,k_{v}+\Delta k\} & \text{if } q_{tk_{uv}} > q_{tk_{v}}
            \end{cases}$
        \STATE Make the probability values at $k_{sup}$ equal to 0 and renormalize the bin probabilities as \\
        $q^{'}_{tk} = \begin{cases}
                0 & \text{if } k\in k_{sup} \\
                \frac{q_{tk}}{1-\underset{k\in{k_{sup}}}{\sum}q_{tk}}  & \text{if } k \notin k_{sup}
            \end{cases}$
        \ENDIF
    \ENDFOR
  \end{algorithmic}
\end{algorithm}

During testing, we observed that there are a few instances where the predicted distribution $q(y_{t}|X)$ exhibits two simultaneous peaks $-$ one at $k=1$, i.e. at the unvoiced bin and another at a voiced bin within the range $k \in [k_{v1},k_{v2}]$. This can lead to an incorrect expected value computed using eq.~\ref{eq:expected_y}, as the presence of these simultaneous peaks may skew the predicted point estimate towards an intermediate value that does not accurately reflect the true pitch. To address this, we apply a post-processing pruning algorithm $P$ that updates $q(y_{t}|X)$ by suppressing the less probable of the two peaks, which is detailed in Algorithm~\ref{pruning}. This is pictorially depicted in Fig.~\ref{prune-example}. It is important to note that pruning is applied only when two peaks occur simultaneously—one at the unvoiced bin and another at a voiced bin. Pruning is not performed when multiple peaks are present solely within the voiced bin range. 

Further, we calculate the predicted standard deviation from $q(y_t|X)$ at time frame $t$ by
\begin{equation}
    \hat{\sigma}_{t} = \sqrt{\mathbb{E}_{\hat{y}\sim q(y_{t}|X)}[(\hat{y} - \hat{y}_{t})^2]}
    \label{eq:pred_std_dev}
\end{equation}

where $\hat{\sigma}_t$ is the uncertainty estimate. At this point, we make an assumption that after training the model $f_{\theta}$ using M1, the predicted $\hat{\sigma}$ does not reflect the deviation of the mean $\hat{y}$ from the true value $y$, which is substantiated in Section~\ref{results}. To address this issue, we propose an alternative method, which is described in the following section. 

\subsection{Histogram loss with dynamic standard deviation \texorpdfstring{$\sigma$}{sigma} (M2)} \label{M2}
This method is almost similar to M1 but with a slight modification. In this method, the standard deviation of the target distribution is no longer equal to bin width $b_w$ as in M1; instead, it is dynamically adjusted, as explained below.

With $X$ as the input, for a particular time frame $t$, the base model $f_{\theta}$ predicts the predicted probability distribution $q(y_{t}|X) = f_{\theta}(X) = (q_{t1},q_{t2},...,q_{tK}) $. From this, we calculate the mean $\hat{y}_{t}$ using eq.~\ref{eq:expected_y}. We consider a target distribution $p(y_{t}|X)$ as a Gaussian distribution with mean $y_{t}$, but instead of a fixed standard deviation $\sigma_t$ equal to the bin width $b_w$, we define it dynamically based on the prediction error between $\hat{y}_{t}$ and $y_{t}$, i.e., $\sigma_t = \text{sg}[|y_t-\hat{y}_t|]$, where $\text{sg}[\cdot]$ represents the stop gradient\footnote{The “stop gradient” notation $\text{sg}[\cdot]$ indicates the argument is treated as fixed when computing a
gradient.}. Therefore, the target distribution becomes $p(y_{t}|X)=\mathcal{N}(y_{t},\text{sg}[(y_{t}-\hat{y}_{t})^2])$. Notably, while the bin weights $p_{tk}$ for each bin $k$ have previously been precomputed, they are now computed in real-time, as the standard deviation $\sigma_{t}$ depends on the predicted mean $\hat{y}_{t}$. During training, the base model parameters $\theta$ are updated using eq.~\ref{gd}, with the loss $\mathcal{L}_{wHL}$ (in eq.~\ref{whl}) calculated by using the real-time bin weights $p_{tk}$ for each bin $k$. We train the base model $f_{\theta}$ for $E_2$ epochs. After training the base model, we predict the uncertainty estimates $\hat{\sigma}_t$ for each time frame using eq.~\ref{eq:pred_std_dev}. A point to note is that since this method explicitly models the standard deviation, we observed that it inherently mitigates the occurrence of multiple peaks at unvoiced and voiced bins, thereby eliminating the need for pruning or additional post-processing, detailed in Section~\ref{ablation}. Instead of assigning an arbitrary value to unvoiced frames, i.e., 5 bins below $g(51.91)$, a more principled approach is to treat voiced/unvoiced detection as a classification task and log-frequency prediction for voiced frames as a regression problem, analogous to a full Bayesian setting as explained below.

\subsection{Histogram loss with dynamic standard deviation \texorpdfstring{$\sigma$}{sigma} in full Bayesian setting (M3)}
Consider a dataset $D = \{(X_i,y_i,v_i)\}_{i=1}^{I}$, where $y_i$ is a vector of dimension $T$, consisting of log-frequency values for voiced frames computed using eq.~\ref{log_freq_fn}. Since the log-frequency values are only computed for voiced frames, we restrict the support range to the voiced interval $[0,4]$, which is uniformly partitioned into $K=385$ bins. Also, $v_i$ is a voicing vector of dimension $T$, where $v_{it} = 1$ for voiced frames and $v_{it} = 0$ for unvoiced frames. 

For a given input $X$, the model predicts the voicing probability $q(v_t|X)$ (voicing detection) for each time frame $t$ and, for voiced frames, a predictive histogram over log-frequency bins $q(y_t|v_t=1,X)$ (pitch detection). The voicing probabilities are trained using a weighted binary cross-entropy loss defined as:

\begin{equation}
\begin{aligned}
    \mathcal{L}_{wBCE} = - \sum_{i,t} w_{c_{it}} [v_{it} \ln q(v_{it}|X_i) \\ 
    \quad + (1-v_{it}) \log (1-q(v_{it}|X_i))]
\end{aligned}
\label{bce}
\end{equation}


where $w_{c_{it}}$ are the weights of the voiced and unvoiced classes computed using eq.~\ref{wcs}. For the voiced frames, the predictive histogram over log-frequency bins is trained using the histogram loss $\mathcal{L}_{HL}$ (in eq.~\ref{hil}) calculated using real-time bin weights $p_{k}$ for each bin $k$ as calculated in M2 method explained in Section~\ref{M2}. The total loss for training the model is a weighted combination of the above losses and is defined as:
\begin{equation}
    \mathcal{L}_{B} = \mathcal{L}_{wBCE} + \lambda \mathcal{L}_{HL}
    \label{final_mle}
\end{equation}
where $\lambda = 0.6$ is the scaling factor.

\begin{figure}
    \centering
    \includegraphics[width=\columnwidth]{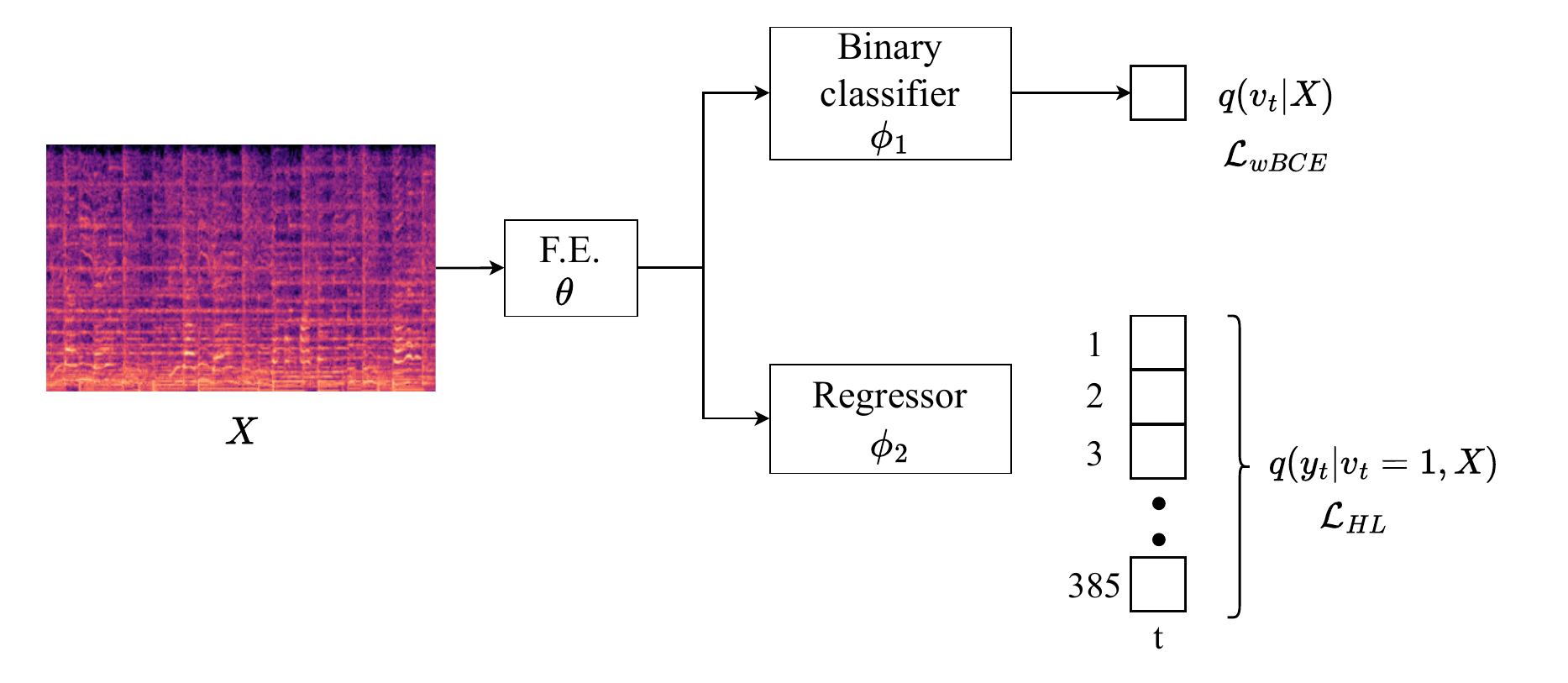}
    \caption{Here, $\theta$, $\phi_1$, and $\phi_2$ represent the parameters of the feature extractor layer, classifier layer, and regressor layer, respectively. At a particular time frame $t$, if $v_t = 0$, only $\mathcal{L}_{BCE}$ is calculated, whereas, if $v_t = 1$, then both $\mathcal{L}_{BCE}$ and $\mathcal{L}_{HL}$ are calculated.}
    \label{mle_model}
\end{figure}

Consider a base model as in Fig.~\ref{mle_model} where $\theta$ are the parameters of the feature extractor layers, $\phi_1$ are the parameters of the classifier layer, and $\phi_2$ are the parameters of the regression layer. During training, the model parameters $\theta$, $\phi_1$, and $\phi_2$ are updated using the gradient descent algorithm as:
\begin{equation}
    [\theta,\phi_1,\phi_2] \leftarrow [\theta,\phi_1,\phi_2] - \alpha \nabla_{[\theta,\phi_1,\phi_2]} \mathcal{L}_{B}(f_{[\theta,\phi_1,\phi_2]})
\end{equation}
where $\alpha \in \mathbb{R}^+$ is the learning rate. We train the model $f_{[\theta,\phi_1,\phi_2]}$ for $E_3$ epochs. After training the model, we predict the uncertainty estimates $\hat{\sigma}$ for the voiced frames using eq.~\ref{eq:pred_std_dev}. 

\begin{table*}[t]
\caption{Performance metrics with our base models across all the proposed methods and other baseline models. All the models are trained on the train data $D$ and evaluated on the three test datasets. Here, Cls and Reg stand for classification and regression approaches, respectively, for the melody estimation problem. Here, $P(\cdot)$ represents the results after applying the pruning algorithm. All values are in percentages. The $\pm$ values are the bootstrap 95\% confidence intervals.}
\label{cls-vs-reg}
\centering
\resizebox{\textwidth}{!}{
  \begin{tabular}{|*{11}{c|}} \cline{1-11} 
  \hline
  \multicolumn{1}{|c|}{\multirow{2}{*}{\textbf{Experiments}}} & \multicolumn{1}{c|}{\multirow{2}{*}{\textbf{Approach}}} & \multicolumn{3}{c|}{\textbf{ADC2004}} & \multicolumn{3}{c|}{\textbf{MIREX05}} & \multicolumn{3}{c|}{\textbf{HAR}}\\ \cline{3-11}
  \multicolumn{1}{|c|}{} & \multicolumn{1}{c|}{} & \textbf{RPA} & \textbf{RCA}  & \textbf{OA} & \textbf{RPA} & \textbf{RCA}  & \textbf{OA} & \textbf{RPA} & \textbf{RCA} & \textbf{OA} \\
  \hline  
  Patch-based CNN\cite{patch-basedcnn}  & Cls &  78.03 $\pm$ 3.12 & 79.82 $\pm$ 3.09 & 80.12 $\pm$ 3.24 & 76.55 $\pm$ 4.21 & 83.13 $\pm$ 4.12  & 83.56 $\pm$ 4.90 & 70.02 $\pm$ 3.21 & 71.45 $\pm$ 3.30 & 69.43 $\pm$ 3.34\\
  \hline
  NMF-CRNN\cite{nmf-crnn}  & Cls &  78.34 $\pm$ 4.32 & 78.96 $\pm$ 4.20 & 76.27 $\pm$ 5.10 & 78.87 $\pm$ 3.56 & 79.60 $\pm$ 3.80 & 78.15 $\pm$ 4.23 & 69.23 $\pm$ 3.30 & 70.34 $\pm$ 3.33 & 69.40 $\pm$ 3.10\\
  \hline
  Attention Network\cite{attention} & Cls & 77.03 $\pm$ 2.90  & 78.05 $\pm$ 2.31 & 79.46 $\pm$ 3.67 & 79.81 $\pm$ 3.78 & 79.85 $\pm$ 3.60 & 86.33 $\pm$ 3.90 & 69.56 $\pm$ 2.89 & 70.17 $\pm$ 2.78 & 69.80 $\pm$ 2.95\\
  \hline
  SegNet\cite{en-decoder} & Cls & 82.45 $\pm$ 3.45 & 83.90 $\pm$ 3.53 & 80.60 $\pm$ 3.78 & 79.48 $\pm$ 4.32 & 80.34 $\pm$ 4.50 & 79.29 $\pm$ 3.78 & 70.43 $\pm$ 3.23 & 71.23 $\pm$ 3.50 & 67.36 $\pm$ 4.02\\
  \hline
  HKDSME\cite{hkdsme} & Cls & 82.24 $\pm$ 2.76 & 83.16 $\pm$ 2.90 & 82.45 $\pm$ 3.10 & 83.45 $\pm$ 3.15 & 83.49 $\pm$ 3.60 & 84.19 $\pm$ 3.40 & 79.23 $\pm$ 3.90 & 80.12 $\pm$ 3.85 & 79.02 $\pm$ 3.24 \\
  \hline
  MTANET\cite{mtanet} & Cls & 81.56 $\pm$ 2.54 & 82.19 $\pm$ 2.60 & 82.10 $\pm$ 2.10 & 84.10 $\pm$ 2.78 & 84.34 $\pm$ 2.85 & 82.40 $\pm$ 2.90 & 80.45 $\pm$ 2.74 & 81.23 $\pm$ 2.90 & 79.32 $\pm$ 3.10 \\
  \hline
  HANET\cite{hanet} & Cls & 84.56 $\pm$ 2.15 & 85.04 $\pm$ 2.30 & 84.22 $\pm$ 2.78 & 84.50 $\pm$ 3.01 & 84.89 $\pm$ 3.10 & 83.90 $\pm$ 3.56 & 86.45 $\pm$ 2.98 & 86.98 $\pm$ 2.78 & 86.10 $\pm$ 2.01\\
  \hline
  ToNet\cite{tonet} & Cls & 80.34 $\pm$ 3.03 & 81.32 $\pm$ 3.10 & 81.19 $\pm$ 3.45 & 81.80 $\pm$ 2.56 & 82.23 $\pm$ 2.89 & 81.90 $\pm$ 3.01 & 78.90 $\pm$ 3.22 & 79.23 $\pm$ 3.40 & 78.45 $\pm$ 3.55\\
  \hline
  \hline
  M-MSE & Reg & 21.66 $\pm$ 5.06 & 22.67 $\pm$ 5.10 & 20.42 $\pm$ 4.12 & 25.74 $\pm$ 4.89 & 26.70 $\pm$ 4.90 & 24.15 $\pm$ 5.32 & 45.98 $\pm$ 4.78 & 46.19 $\pm$ 4.82 & 46.27 $\pm$ 4.43 \\
  \hline
  M-NLL\cite{seitzer} & Reg & 68.08 $\pm$ 4.23 & 68.74 $\pm$ 4.67 & 59.20 $\pm$ 4.65 & 68.82 $\pm$ 4.90 & 69.77 $\pm$ 4.98 & 57.63 $\pm$ 5.01 & 95.69 $\pm$ 4.67 & 95.85 $\pm$ 4.89 & 89.75 $\pm$ 5.22\\
  \hline
  \hline
  \textbf{M1} & Reg & 84.04 $\pm$ 2.89 & 84.25 $\pm$ 2.85 & 84.41 $\pm$ 3.14 & 85.65 $\pm$ 2.14 & 85.80 $\pm$ 2.11 & 91.20 $\pm$ 1.11 & 98.27 $\pm$ 0.06 & 98.31 $\pm$ 0.20 & 98.78 $\pm$ 0.03\\
  \hline
  \textbf{$P$(M1)} & Reg & 85.99 $\pm$ 2.62 & 86.05 $\pm$ 2.60 & 86.55 $\pm$ 2.69 & 89.46 $\pm$ 1.99 & 89.46 $\pm$ 1.99 & 94.32 $\pm$ 0.92 & 98.89 $\pm$ 0.07 & 98.90 $\pm$ 0.18 & 99.28 $\pm$ 0.03\\
  \hline
  \textbf{M2} & Reg & 87.06 $\pm$ 2.60 & 87.16 $\pm$ 2.57 &  86.81 $\pm$ 2.83 & 89.51 $\pm$ 1.63 & 89.54 $\pm$ 1.61 & 93.67 $\pm$ 0.85 & 98.91 $\pm$ 0.08 & 98.95 $\pm$ 0.07 & 99.20 $\pm$ 0.03 \\
  \hline
  \textbf{M3} & Reg & \textbf{87.71 $\pm$ 2.08} & \textbf{87.88 $\pm$ 2.10} & \textbf{86.82 $\pm$ 2.56} & \textbf{96.10 $\pm$ 1.08} & \textbf{96.11 $\pm$ 1.07} & \textbf{97.38 $\pm$ 0.76} & \textbf{99.48 $\pm$ 0.05} & \textbf{99.49 $\pm$ 0.05} &\textbf{99.60 $\pm$ 0.03} \\
  \hline
  \end{tabular}}
\end{table*}

\begin{table}[t]
\caption{NLL values calculated with our methods and the other baseline regression method on the three test datasets.}
\centering
  \resizebox{0.8\columnwidth}{!}{
  \begin{tabular}{|*{4}{c|}} \cline{1-4}
  \textbf{Experiments} & \textbf{ADC2004} & \textbf{MIREX05} & \textbf{HAR} \\
  \hline
  M-NLL &  3.36   &  0.89   &  1.32   \\
  \hline
  M1 &   24.29  &   10.21    &   0.33\\
  \hline
  M2 & 22.31   &   11.48  &    0.49  \\
  \hline
  \textbf{M3} &  \textbf{-2.82} &   \textbf{-3.53} & \textbf{-3.91} \\
  \hline 
  \end{tabular}}
  \label{nll-metric}
\end{table}

\section{Experiments} \label{exps}
\subsection{Data}
For the melody estimation task, we train on two datasets $D$ $-$ the first is MIR1K{\footnote{\url{http://mirlab.org/dataset/public/}}} consisting of 1000 Chinese karaoke clips of 2.2 hours. The second is a subset of the HAR{\footnote{\label{note2}\url{https://zenodo.org/record/8252222}}} dataset consisting of 259 audio recordings of 2.6 hours from one teacher. No data augmentation is applied. We have tested the performance of the model on the three test datasets $-$ ADC2004{\footnote{\label{note1}\url{http://labrosa.ee.columbia.edu/projects/melody/}}}, MIREX05\footref{note1}, and the remaining recordings from the other teacher in the HAR\footref{note2} dataset. The proposed model is only trained for singing voice melody, so we have selected only those test samples that contained melody sung by humans. As a result, 12 clips in ADC2004, 9 clips in MIREX05, and 264 clips in HAR are selected. Since we divide the audios into 1-second chunks, we have 17348 audio chunks in train data $D$; and 98, 198, and 9622 audio chunks in ADC2004, MIREX05, and HAR, respectively.

\subsection{Experiment Setting}
In this paper, we employ a basic CRNN model as the base model. For M1 and M2, the base model consists of 4 ResNet blocks with $f = [32,64,128,256]$ filters followed by a TimeDistributed Dense layer with $K=435$ nodes with softmax activation. Each ResNet block includes: a $1\times1$ convolutional layer with $f$ number of channels with Batch Normalization (BN) and a LeakyReLU activation with a slope of 0.01, followed by two $3\times 3$ convolutional layers with $f$ channels each with BN and LeakyReLU activation, and a final $1\times 1$ convolutional layer with $f$ channels with BN. A shortcut connection is added after the first $1\times1$ convolution, and the summed output is passed through LeakyReLU and a $1\times4$ MaxPooling layer. For M3, the same 4 ResNet blocks are followed by two branches: a Dense layer with a single node with sigmoid activation for voicing detection, and a Dense layer with $K=385$ nodes with softmax activation for voiced pitch detection. All models are trained for 100 epochs each, i.e., $E_1=E_2=E_3=100$. 

We compare the performance of our proposed methods with the baseline experiments. To maintain a valid comparison, we keep the same train and test data across all the baseline experiments. We categorize the experiments into three categories: melody estimation, performance with NLL, and uncertainty estimation. We explain the experiments as follows:
\begin{enumerate}
    \item \textbf{Melody estimation:} We train the base models across all methods, on the train data $D$ for 100 epochs by using a learning rate of $\alpha = 1\times10^{-5}$. The trained base models are used to evaluate the performance on the three test datasets. We compare the performance of our methods with the following:
    \begin{itemize}
        \item Existing non-regression baselines that treat melody estimation as a classification problem. This includes Patch-based CNN~\cite{patch-basedcnn}, NMF-CRNN~\cite{nmf-crnn}, Attention Network~\cite{attention}, SegNet~\cite{en-decoder}, HKDSME~\cite{hkdsme}, MTANET~\cite{mtanet}, HANET~\cite{hanet}, and ToNet~\cite{tonet}. We have obtained the results of these experiments on the audios in the three test datasets by downloading their online source codes and compiling the results on our dataset configuration.
        \item Base model in M1 trained with existing losses for regression tasks. The model consists of 4 ResNet blocks followed by an output layer that varies depending on the chosen loss function. The models are trained on train data $D$ and tested on three test datasets. The experiments are defined as: 
        \begin{enumerate}
            \item M-MSE: The output layer is a Dense layer with a single node and linear activation function. This model is trained for 100 epochs using mean squared error as the loss function.
            \item M-NLL: The output layer consists of two branches $-$ one predicting the mean through a Dense layer with a single node and linear activation, and the other predicting the variance through a Dense layer with a single node and softplus activation. This model is trained for 250 epochs by using negative log-likelihood loss~\cite{seitzer}. A point to note here is that this experiment required more epochs to reach convergence, whereas the other methods converged in 100 epochs.
        \end{enumerate}
    \end{itemize}
    The performance metrics considered are raw pitch accuracy (RPA), raw chroma accuracy (RCA), and overall accuracy (OA). All these metrics are computed by using a standard \textit{mir-eval}~\cite{mir_eval} library with a pitch detection tolerance of 50 cents.

    \item \textbf{Performance with NLL:} To measure how well the predicted distribution matches the target distribution of dataset $D$, we use the negative log-likelihood (NLL) as the evaluation metric, defined as:
    \begin{equation}
        NLL(D) = \frac{1}{2|D|}\sum_{i,t} \ln(2\pi \hat{\sigma}^2_{it}) + \frac{(y_{it}-\hat{y}_{it})^2}{\hat{\sigma}_{it}^2}
    \end{equation}
    where a lower NLL value indicates better model performance. We compare the NLL values calculated from our proposed methods with those from M-NLL on the three test datasets.

    \item \textbf{Uncertainty estimation:} This experiment is further divided into the following-
    
    \begin{enumerate}
        \item \textbf{Comparison with regression baselines:} After training the base models, we compare the uncertainty estimates $\hat{\sigma}$ obtained by our proposed methods with those from M-NLL. To evaluate how well the predicted $\hat{\sigma}$ reflects the deviation $|y - \hat{y}|$, we plot $\hat{\sigma}$ against $|y - \hat{y}|$ for all the methods. 

       \item \textbf{Comparison with classification-based confidence baseline (CBC):} To assess whether the proposed regression-based uncertainty estimation provides any advantage over the classification-based confidence method~\cite{saxena2024}, we conduct a mistake detection experiment. The objective of this experiment is to examine whether frames assigned lower confidence values correspond to incorrect pitch predictions. Specifically, we rank all frames in ascending order of confidence and evaluate how well the least confident frames align with actual prediction errors. For the regression-based method, confidence is defined to be inversely proportional to the predicted uncertainty, referred to as regression-based confidence (RBC). The performance is quantified using the F1-score for detecting errors among the $U$ least confident frames, where $U=\{10,20,30\}$. A time frame is considered incorrect if the predicted pitch deviates by more than 50 cents from the ground truth.

    \end{enumerate}

\end{enumerate}


\begin{figure*}[t]
    \centering
    \includegraphics[width=18cm,height=15cm]{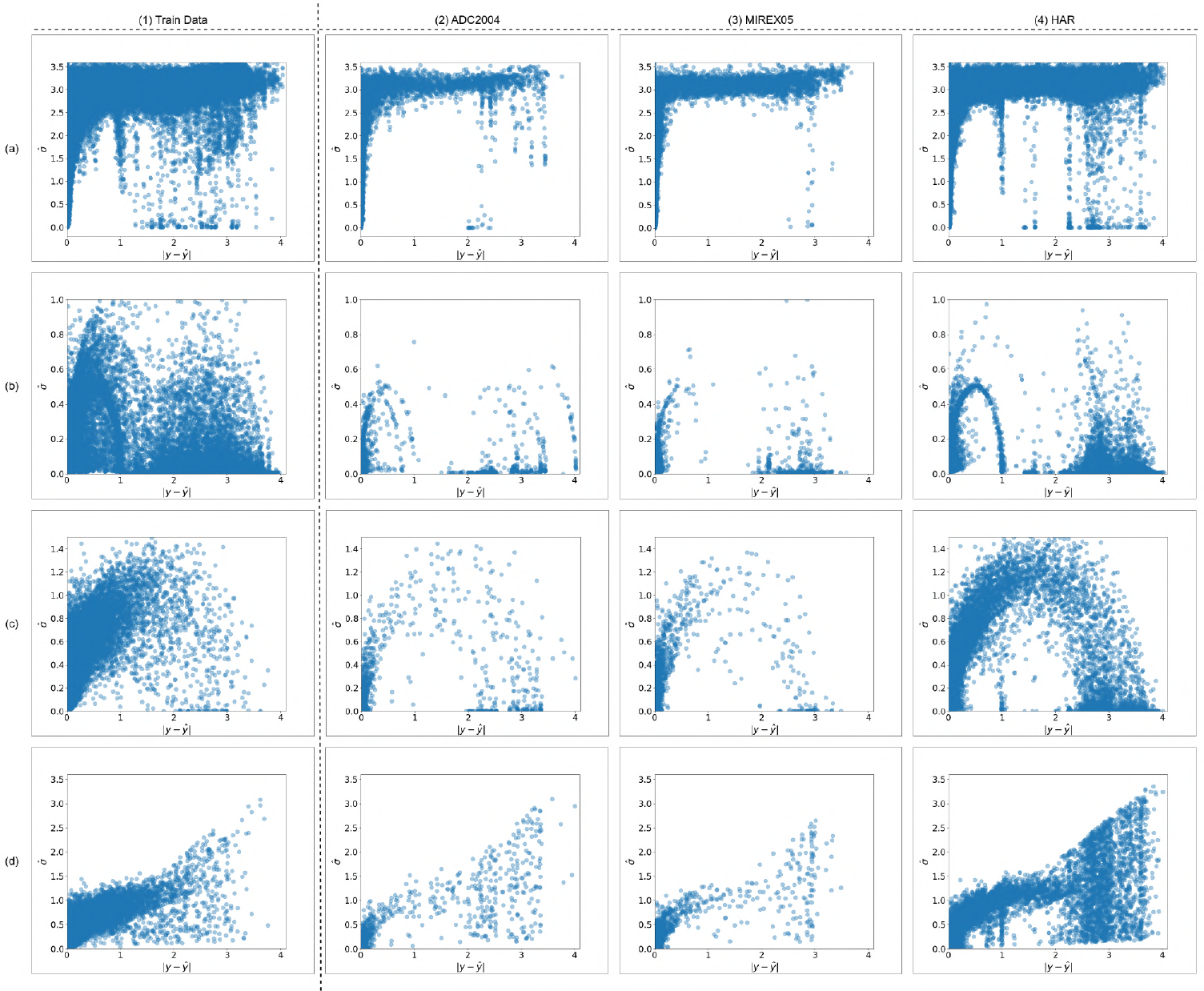}
    \caption{Uncertainty estimates $\hat{\sigma}$ vs prediction error $|y-\hat{y}|$ obtained from (a) M-NLL model, and models trained using (b) M1, (c) M2, and (d) M3, on train as well as three test datasets. Here column (1) represents the train data, and the rest of the columns (2)-(4) represent a different test dataset, while each row (a)-(d) corresponds to a regression-based method.  Plots (a), (b), and (c) include both unvoiced and voiced frames, while plot (d) only considers voiced frames, as voiced pitch detection in M3 is treated as a regression task.}
    \label{diff_vs_dev}
\end{figure*}

\section{Results}
\label{results}
Table~\ref{cls-vs-reg} depicts the comparison of melody estimation performance between classification- and regression-based approaches across the three test datasets. All values reported in Table~\ref{cls-vs-reg} are presented with 95\% bootstrap confidence intervals, computed over 1000 resamples (with replacement), providing a measure of the statistical confidence of each metric (explanation in Section~\ref{bootstrapping}). We observe that the proposed regression-based methods — M1, M2, and M3- consistently outperform the classification-based baseline methods. The suboptimal performance of the classification-based methods can be attributed to class imbalance in the discretized pitch bins, which can lead to overall performance degradation.

Amongst the proposed regression-based approaches, M1 demonstrates a notable improvement. Applying the pruning algorithm to M1, denoted by $P$(M1), further enhances the performance by effectively mitigating the errors caused by simultaneous peaks in the unvoiced and voiced bins, as discussed in Section~\ref{M1}. M2 builds upon M1 by refining the modeling process, where the standard deviation of the target distribution is explicitly modeled to reflect the prediction error, thereby achieving better accuracy. However, the best performance is observed with M3, which consistently outperforms all other proposed methods. This highlights that the Bayesian approach to melody estimation is a more effective and principled way to capture the continuous nature of melody. 

The results indicate that the regression-based baselines exhibit inferior performance as compared to the proposed methods. The poor performance of M-MSE can be attributed to its inherent limitation of treating melody estimation as a pointwise regression problem. While M-NLL performs better than M-MSE, it still falls short of the proposed methods. Although M-NLL models the target distribution as a Gaussian, it imposes a fixed distribution shape that may not align with the true underlying data, leading to suboptimal performance. The relatively narrow confidence intervals for the proposed methods, particularly M3, indicate that the performance improvements are consistent and not due to random variations in the data. Notably, the HAR dataset achieves the highest performance across all the proposed methods, due to its clean, studio-recorded audio with minimal noise, which facilitates more accurate melody estimation.


\begin{table*}[t]
\caption{F1-scores for mistake detection on the $U$ least confident from CBC and RBC methods, respectively, across three test datasets. All values are in percentages.}
\label{mistakedetection}
\centering
\resizebox{0.8\textwidth}{!}{
  \begin{tabular}{|*{10}{c|}} \cline{1-10} 
  \multicolumn{1}{|c|}{\multirow{2}{*}{\textbf{Experiments}}}  & \multicolumn{3}{c|}{\textbf{ADC2004}} & \multicolumn{3}{c|}{\textbf{MIREX05}} & \multicolumn{3}{c|}{\textbf{HAR}} \\ \cline{2-10}
  \multicolumn{1}{|c|}{} & \textbf{U=10} & \textbf{U=20}  & \textbf{U=30} & \textbf{U=10} & \textbf{U=20}  & \textbf{U=30} & \textbf{U=10} & \textbf{U=20} & \textbf{U=30}\\
  \hline  
  CBC & 74.12 & 74.98 & 75.12 & 72.90 & 72.02 & 72.20 & 75.75 & 75.78 & 76.80 \\
  \hline  
  RBC & 82.19 & 82.78 & 81.86 & 82.81 & 82.82 & 83.28 & 82.86 & 83.10 & 83.90\\
  \hline  
  \end{tabular}}
\end{table*}

Table~\ref{nll-metric} presents the calculated NLL values for our proposed methods and the baseline model M-NLL across three test datasets. The results show that M3 outperforms both the baseline and other methods, achieving a better alignment between predicted and target distributions. Additionally, we observe a trend in NLL values that reflects the relationship between uncertainty estimates and prediction error. The higher NLL values for M1 and M2 indicate that their uncertainty estimates ($\hat{\sigma}$) are often too small for large deviations, leading to a poor fit. M3 achieves the lowest NLL values, indicating that its uncertainty estimates are better correlated with the prediction deviations.

Fig.~\ref{diff_vs_dev} depicts the comparison of the uncertainty estimates $\hat{\sigma}$ obtained from our proposed methods with those from M-NLL.  In Fig.~\ref{diff_vs_dev}(a)(1)-(4), we observe that when the M-NLL model is trained using the negative log-likelihood loss, the estimated $\hat{\sigma}$ remains high even for small prediction errors across all datasets. However, the low value of estimated $\hat{\sigma}$ for larger prediction errors is predominantly observed in the train data (Fig.~\ref{diff_vs_dev}(a)(1)) and the HAR test data (Fig.~\ref{diff_vs_dev}(a)(4)). This indicates that the M-NLL model struggles to correlate the uncertainty estimates $\hat{\sigma}$ with the prediction errors across different datasets. Fig.~\ref{diff_vs_dev}(b)(1)-(4) shows the uncertainty estimates $\hat{\sigma}$ obtained from the M1 method. For low prediction errors, $\hat{\sigma}$ values are lower compared to the M-NLL method, with the majority of values correlating well with the prediction error. However, for larger prediction errors, a significant number of values exhibit low $\hat{\sigma}$, indicating poor correlation between the uncertainty estimates and the actual prediction error. This issue is more pronounced compared to the M-NLL method across all datasets. Interestingly, despite M1 demonstrating good melody estimation performance (as shown in Table~\ref{cls-vs-reg}), its uncertainty estimates do not consistently correlate with the prediction error. Fig.~\ref{diff_vs_dev}(c)(1)-(4) illustrates the uncertainty estimates obtained from the M2 method, which shows an improvement over M1. We observe that in Fig.~\ref{diff_vs_dev}(c)(1)-(4), $\hat{\sigma}$ has started to correlate with low prediction errors.  Additionally, in the Fig.~\ref{diff_vs_dev}(c)(1)-(3), M2 performs better than M1 as $\hat{\sigma}$ now takes larger values for larger prediction errors. In Fig.~\ref{diff_vs_dev}(c)(4), M2 also demonstrates an improvement over M1, with more $\hat{\sigma}$ values correlating with large prediction errors. However, some instances remain where $\hat{\sigma}$ does not fully correlate with the prediction error. Fig.~\ref{diff_vs_dev}(d)(1)-(4) presents the uncertainty estimates obtained from the M3 method, which outperforms all the proposed methods. We observe that $\hat{\sigma}$ now correlates well with the prediction error, even for the large prediction errors. The number of uncorrelated samples is significantly reduced. 

Table~\ref{mistakedetection} compares the F1-scores on the $U$ least confident voiced frames obtained from the CBC and RBC methods, respectively. We observe that the F1-scores achieved by our method RBC consistently surpass those obtained by the CBC method across all values of $U$. This justifies that the confidence estimates obtained from the RBC method are more reliable indicators of actual prediction errors than the confidence scores produced by the CBC model.


Fig. S1 (Supplementary Material) shows the predicted melody, and the corresponding uncertainty estimates $\hat{\sigma}$ for a typical audio sample from the three test datasets using the proposed methods — M1, M2, and M3. Ideally, if the predicted $\hat{\sigma}$ correlates well with the prediction error, the ground truth melody should lie within the uncertainty bounds around the predicted melody. In Fig.~\ref{sigma_melody_plots}(a)(1)-(3), which corresponds to method M1 across all the test datasets, we observe that the uncertainty estimates $\hat{\sigma}$ from M1 do not reflect the prediction error, leading to instances where the ground truth melody falls outside the uncertainty bounds around the predicted melody, particularly at incorrect melody predictions. In Fig.~\ref{sigma_melody_plots}(b)(1)-(3), the accuracy of the predicted melody improves with method M2 compared to M1, leading to better uncertainty estimates $\hat{\sigma}$ that begin to correlate with the prediction error. Fig.~\ref{sigma_melody_plots}(c)(1)-(3), the uncertainty estimates from M3 exhibit a better correlation with the prediction error while also achieving the highest accuracy in melody estimation. Notably, M3 estimates uncertainty only for voiced frames, as voiced pitch detection is treated as a regression task.

\begin{table*}[t]
\caption{Ablation study of the M1 and M2 methods on the three test datasets. Here $P(\cdot)$ represents pruning. All values are in percentages.}
\label{ablation-hil}
\centering
\resizebox{0.8\textwidth}{!}{
  \begin{tabular}{|*{10}{c|}} \cline{1-10} 
  \multicolumn{1}{|c|}{\multirow{2}{*}{\textbf{Experiments}}}  & \multicolumn{3}{c|}{\textbf{ADC2004}} & \multicolumn{3}{c|}{\textbf{MIREX05}} & \multicolumn{3}{c|}{\textbf{HAR}} \\ \cline{2-10}
  \multicolumn{1}{|c|}{} & \textbf{RPA} & \textbf{RCA}  & \textbf{OA} & \textbf{RPA} & \textbf{RCA}  & \textbf{OA} & \textbf{RPA} & \textbf{RCA} & \textbf{OA}\\
  \hline  
  HL-M1  & 78.37 & 79.25 & 72.28 & 75.41 & 75.94 & 74.55 & 95.45 & 95.63 & 95.85 \\
  \hline  
  $P$(HL-M1)  & 79.37 & 79.55 & 72.56 & 76.14 & 76.54 & 75.85 & 96.25 & 96.83 & 96.15 \\
  \hline  
  HL-M2  & 81.20  & 81.65 & 76.32 & 79.12 & 80.43 & 79.72 & 96.89 & 97.12 & 96.33 \\
  \hline
  \textbf{M1}  & 84.04 & 84.25 & 84.41 & 85.65 & 85.80 & 91.20 & 98.27 & 98.31 & 98.78 \\
  \hline
  \textbf{$P$(M1)} &85.99 & 86.05 & 86.55 & 89.46 & 89.46 & 94.32 & 98.89 & 98.90 & 99.28\\
  \hline
  \textbf{M2} & 87.06 & 87.16 &  86.81 & 89.51 & 89.54 & 93.67 & 98.91 & 98.95 & 99.20\\
  \hline
  $P$(M2) & 87.06 & 87.16 &  86.82 & 89.50 & 89.54 & 93.67 & 98.91 & 98.94 & 99.19\\
  \hline
  \end{tabular}}
\end{table*}

\begin{table*}[t]
\caption{Ablation study of the M3 method on the three test datasets. All values are in percentages.}
\label{ablation-m3}
\centering
\resizebox{0.8\textwidth}{!}{
  \begin{tabular}{|*{10}{c|}} \cline{1-10} 
  \multicolumn{1}{|c|}{\multirow{2}{*}{\textbf{Experiments}}}  & \multicolumn{3}{c|}{\textbf{ADC2004}} & \multicolumn{3}{c|}{\textbf{MIREX05}} & \multicolumn{3}{c|}{\textbf{HAR}} \\ \cline{2-10}
  \multicolumn{1}{|c|}{} & \textbf{RPA} & \textbf{RCA}  & \textbf{OA} & \textbf{RPA} & \textbf{RCA}  & \textbf{OA} & \textbf{RPA} & \textbf{RCA} & \textbf{OA}\\
  \hline  
  M3-MSE & 40.25 & 41.65 & 41.14 & 49.34 & 49.44 & 49.13 & 63.12 & 63.13 & 62.89 \\
  \hline  
  M3-NLL & 71.78 & 72.67 & 71.17 & 75.45 & 74.45 & 74.90 & 80.23 & 81.65 & 81.45 \\
  \hline  
  \textbf{M3}  & \textbf{87.71} & \textbf{87.88} & \textbf{86.82} & \textbf{96.10} & \textbf{96.11} & \textbf{97.38} & \textbf{99.48} & \textbf{99.49} & \textbf{99.60} \\
  \hline  
  \end{tabular}}
\end{table*}

\section{Ablation studies}
\label{ablation}
We perform the following ablation experiments:
\begin{enumerate}
    \item HL-M1: This experiment is identical to M1, with model trained using histogram loss (eq.~\ref{hil}) instead of $\mathcal{L}_{wHL}$. We also apply the pruning algorithm to this experiment, denoted by $P$(HL-M1). The model is trained for 100 epochs using the learning rate $\alpha= 1\times 10^{-5}$.
    \item HL-M2: This experiment is identical to M2, with the model trained using histogram loss (eq.~\ref{hil}) instead of $\mathcal{L}_{wHL}$. The model is trained for 100 epochs using the learning rate $\alpha= 1\times 10^{-5}$. 
    \item $P$(M2): After obtaining the trained model from M2 method, we additionally apply pruning algorithm. 
    \item M3-MSE: This experiment is identical to M3, except that the pitch detection output is trained using mean squared error (MSE) loss instead of the histogram loss $\mathcal{L}_{HL}$ in the combined loss $\mathcal{L}_B$ (eq.~\ref{final_mle}). The pitch detection output is a Dense layer with a single node and linear activation. The model is trained for 250 epochs using the learning rate $\alpha= 1\times 10^{-5}$. 
    \item M3-NLL: This experiment is identical to M3, except that the pitch detection output is trained using NLL instead of the histogram loss $\mathcal{L}_{HL}$ in the combined loss $\mathcal{L}_B$ (eq.~\ref{final_mle}). The pitch detection output layer consists of two branches similar to the M-NLL baseline. The model is trained for 250 epochs using the learning rate $\alpha= 1\times 10^{-5}$.
    \item Cent Tolerance Comparison: In this experiment, we assess whether the regression-based M3 method captures fine variations in the melody as compared to classification-based (C1) method~\cite{saxena2024}. Therefore, we compare the performance by both methods on different values of cent tolerance, i.e., $CT=\{12.5,25,37.5,50\}$. 
\end{enumerate}

From Table~\ref{ablation-hil}, we observe that applying pruning enhances the performance of $P$(HL-M1) compared to HL-M1, as explained in Section~\ref{results}. Furthermore, M1 outperforms HL-M1, $P$(M1) surpasses $P$(HL-M1), and M2 demonstrates better performance than HL-M2. This suggests that the performance degradation in the HL-M1 and HL-M2 models may be attributed to the higher occurrence of unvoiced frequency values compared to voiced frequency values. These findings highlight the importance of addressing this imbalance, which M1 and M2 effectively handle. After comparing the performance of $P$(M2) with M2, we observe that pruning is not necessary in M2 as it explicitly models the standard deviation.

Table~\ref{ablation-m3} demonstrates that substituting the histogram loss in the M3 method with mean squared error in the full Bayesian setting (M3-MSE) leads to a substantial performance drop, as MSE yields only point estimates. While replacing the histogram loss with the NLL loss provides some improvement, its unimodal Gaussian assumption limits performance. In contrast, the histogram loss enables the model to capture the full predictive distribution, resulting in more accurate pitch estimation.

Table~\ref{centtolerance} compares the performance of the M3 method with a classification-based baseline~\cite{saxena2024} across different cent tolerance values. We observe that M3 consistently outperforms the classification baseline, even under strict cent tolerance, highlighting that a regression-based approach effectively captures the finer frequency variations in the melody as compared to classification-based method.


\section{Conclusion}
\label{conclusion}
This work presents a new approach to uncertainty estimation that correlates closely with pitch deviation by reformulating melody estimation as a histogram-based regression problem. We design three methods that leverage histogram representations to model pitch over a continuous support range. Among these, the third method, i.e., the Bayesian approach (M3), achieves the best overall performance, providing both improved melody estimation accuracy and uncertainty estimates that strongly correlate with actual prediction errors, thereby enhancing the trustworthiness of melody predictions. Furthermore, estimating uncertainty directly from regression-based formulations offers a principled way to quantify predictive reliability, addressing the limitations of existing classification-based confidence measures.


\bibliographystyle{IEEEtran}
\bibliography{bibliography}

\clearpage
\renewcommand{\thesection}{S\arabic{section}}
\onecolumn
\section*{Supplementary Material\\
for\\
`Uncertainty Quantification in Melody Estimation using Histogram Representation'}

\renewcommand{\thefigure}{S\arabic{figure}}
\setcounter{figure}{0}

\renewcommand{\thetable}{S\arabic{table}}
\setcounter{table}{0}

\renewcommand{\thealgorithm}{S\arabic{algorithm}}
\setcounter{algorithm}{0}

\renewcommand{\thesection}{S\arabic{section}}
\setcounter{section}{0}


\begin{figure}[t]
    \centering
    \includegraphics[width=18cm, height=12.5cm]{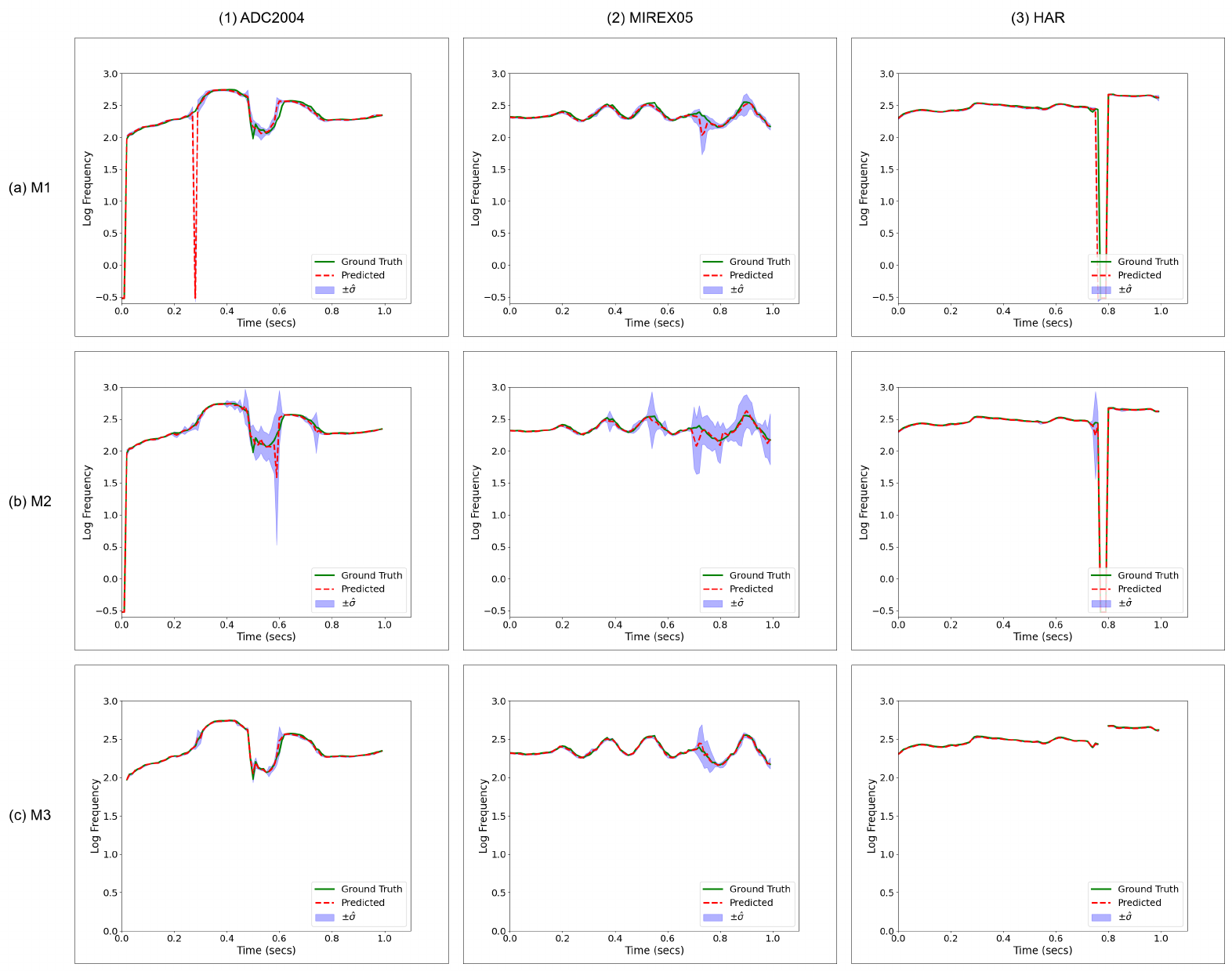}
    \caption{Predicted melody and its corresponding uncertainty estimate $\hat{\sigma}$ for a typical audio sample from the three test datasets using the proposed methods — M1, M2, and M3. Here, columns (1)-(3) represent a different test dataset, while each row (a)-(c) corresponds to a proposed method. The plot displays the ground truth melody (green), the predicted melody (red dashed line), and the uncertainty estimates ($\pm \hat{\sigma}$) around the predictions. In (c), M3 only considers voiced frames, as voiced pitch detection is treated as a regression task.}
    \label{sigma_melody_plots}
\end{figure}

\begin{table*}[t]
\caption{Performance of the M3 method as compared to a classification baseline (C1) on different values of cent tolerance (CT). All values are in percentages.}
\label{centtolerance}
\centering
\resizebox{0.8\textwidth}{!}{
  \begin{tabular}{|c|c|*{12}{c|}} 
  \hline
  \multirow{2}{*}{\textbf{Datasets}} & \multirow{2}{*}{\textbf{Experiments}} 
  & \multicolumn{3}{c|}{\textbf{CT = 12.5}} 
  & \multicolumn{3}{c|}{\textbf{CT = 25}} 
  & \multicolumn{3}{c|}{\textbf{CT = 37.5}} 
  & \multicolumn{3}{c|}{\textbf{CT = 50}} \\ 
  \cline{3-14}
  & & \textbf{RPA} & \textbf{RCA} & \textbf{OA} 
  & \textbf{RPA} & \textbf{RCA} & \textbf{OA} 
  & \textbf{RPA} & \textbf{RCA} & \textbf{OA} 
  & \textbf{RPA} & \textbf{RCA} & \textbf{OA} \\  
  \hline
  \multirow{2}{*}{\textbf{ADC2004}} 
  & C1 & 57.38 & 57.50 & 55.40 & 75.91 & 76.82 & 72.46 & 80.48 & 81.08 & 78.53 & 83.26 & 84.55 & 83.90 \\ \cline{2-14}
  & \textbf{M3} & \textbf{75.27} & \textbf{75.34} & \textbf{78.66} & \textbf{80.04} & \textbf{80.24} & \textbf{80.38} & \textbf{84.38} & \textbf{84.60} & \textbf{84.84} & \textbf{87.71} & \textbf{87.88} & \textbf{86.82} \\
  \hline
  \hline
  \multirow{2}{*}{\textbf{MIREX05}} 
  & C1 & 59.89 & 60.30 & 60.10 & 78.87 & 79.40 & 78.90 & 82.52 & 84.01 & 74.50 & 86.23 & 87.50 & 79.78 \\ \cline{2-14}
  & \textbf{M3} & \textbf{73.85} & \textbf{73.85} & \textbf{83.06} & \textbf{88.47} & \textbf{88.48} & \textbf{92.88} & \textbf{93.24} & \textbf{93.24} & \textbf{95.70} & \textbf{96.10} & \textbf{96.11} & \textbf{97.38} \\
  \hline
  \hline
  \multirow{2}{*}{\textbf{HAR}} 
  & C1 & 61.23 & 62.66 & 62.34 & 68.45 & 69.03 & 68.67 & 75.23 & 76.23 & 75.30 & 79.40 & 80.70 & 79.90 \\ \cline{2-14}
  & \textbf{M3} & \textbf{96.34} & \textbf{96.35} & \textbf{97.24} & \textbf{98.91} & \textbf{98.92} & \textbf{99.20} & \textbf{99.32} & \textbf{99.33} & \textbf{99.50} & \textbf{99.48} & \textbf{99.49} & \textbf{99.60} \\
  \hline
  \end{tabular}}
\end{table*}

\section{Bootstrapping}
\label{bootstrapping}
To estimate the statistical confidence of the reported performance metrics, we employ non-parametric bootstrapping. Consider a test dataset containing $N$ audio clips. Let $M_i$ denote the performance metric (e.g., RPA) computed for clip $n=1,2,...,N$, and $\bar{M}$ represent the mean metric over all clips. To assess the variability of $\bar{M}$ that would be expected if the evaluation were repeated on different samples from the same data distribution, we apply the following bootstrapping procedure:
\begin{enumerate}
    \item \textbf{Bootstrapping: Resampling with Replacement}\\
    We generate $B=1000$ bootstrap samples. For each bootstrap iteration $b=1,2,...,B$, we randomly draw $N$ clips with replacement from the original dataset, allowing some clips to appear multiple times while others may be omitted. The mean performance metric $\bar{M}^{(b)}$ is computed on each resampled set, yielding an empirical distribution $\bar{M}^{(1)}, \bar{M}^{(2)},... \bar{M}^{(1000)}$.

    \item \textbf{Estimating the Confidence Interval}\\
    The bootstrap estimates are sorted, and the $2.5^{th}$ and $97.5^{th}$ percentiles are taken as the bounds of the 95\% confidence interval, denoted as $[CI_{2.5},CI_{97.5}]$.

    \item \textbf{Interpretation}\\
    The 95\% confidence interval indicates the range within which the true mean performance metric is expected to lie in 95\% of repeated experiments conducted under similar conditions. It reflects statistical reliability of the reported performance, rather than uncertainty in individual model predictions.    
\end{enumerate}
\end{document}